\documentclass[apj]{emulateapj}
\usepackage{color}
\usepackage[flushleft]{threeparttable}

\begin{document}
\voffset -2cm

\title{The FRB 121102 host is atypical among nearby FRBs}


\author{Ye Li$^{1, 2}$, Bing Zhang$^{3}$, Kentaro Nagamine$^{4,3,5}$, Jingjing Shi$^{1}$
}

\affil{
$^1$Kavli Institute for Astronomy and Astrophysics, Peking University, Beijing 100871, China \\
$^2$Purple Mountain Observatory, Chinese Academy of Sciences, Nanjing 210008, China \\
$^3$Department of Physics and Astronomy, University of Nevada, Las Vegas, NV 89154, USA \\
$^4$ Theoretical Astrophysics, Department of Earth and Space Science, Graduate School of Science, Osaka University, 1-1 Machikaneyama, Toyonaka, Osaka 560-0043, Japan\\
$^5$Kavli-IPMU (WPI), University of Tokyo, 5-1-5 Kashiwanoha, Kashiwa, Chiba, 277-8583, Japan \\
}

\begin{abstract}

We search for host galaxy candidates of nearby fast radio bursts (FRBs), FRB 180729.J1316+55, FRB 171020, FRB 171213, FRB 180810.J1159+83, and FRB 180814.J0422+73 (the second repeating FRB). 
We compare the absolute magnitudes and the expected host dispersion measure $\rm DM_{host}$ of these candidates with that of the first repeating FRB, FRB 121102, as well as those of long gamma ray bursts (LGRBs) and superluminous supernovae (SLSNe), the proposed progenitor systems of FRB 121102.
We find that while the FRB 121102 host is consistent with those of LGRBs and SLSNe, the nearby FRB host candidates, at least for FRB 180729.J1316+55, FRB 171020, and FRB180814.J0422+73, either have a smaller $\rm DM_{host}$ or are fainter than FRB121102 host, as well as the hosts of LGRBs and SLSNe. 
In order to avoid the uncertainty in estimating $\rm DM_{host}$ due to the line-of-sight effect, we propose a galaxy-group-based method to estimate the electron density in the inter-galactic regions, and hence, $\rm DM_{IGM}$. 
The result strengthens our conclusion. We conclude that the host galaxy of FRB 121102 is atypical, and LGRBs and SLSNe are likely not the progenitor systems of at least most nearby FRB sources. {The recently reported two FRB hosts differ from the host of FRB 121102 and also the host candidates suggested in this paper. This is consistent with the conclusion of our paper and suggests that the FRB hosts are very diverse. }

\end{abstract}

\section{Introduction}
Fast Radio Bursts (FRBs) are bright objects in radio, with durations a few milliseconds
(\citealt{2007Sci...318..777L, 2013Sci...341...53T, 2014ApJ...790..101S, 2016PASA...33...45P}, see \citealt{2018NatAs...2..860L} for a review).
The values of their dispersion measure (DM), an indicator of the electron column density along the line of sight, are much larger than the predicted values from the Milky Way galaxy, so they are expected to be of an extragalactic origin.

The origin of FRBs is highly debated. It is known that at least some FRB sources produce repeating bursts \citep{2016Natur.531..202S,2019Natur.566..235C}. These FRBs are usually explained within the ``intrinsic'' models that invoke young pulsars 
\citep{2016MNRAS.458L..19C,2016ApJ...818...19K, 2016MNRAS.457..232C},
or magnetars
\citep{2017ApJ...841...14M, 2017ApJ...843L..26B, 2017ApJ...839L...3K,2019MNRAS.485.4091M},
with the ultimate energy coming either from the spindown power or the magnetic power of a neutron star. 
Alternatively, some ``extrinsic'' models invoking the kinetic energy of an external source (e.g. the so-called ``cosmic comb'' model, \citealt{2017ApJ...836L..32Z, 2018ApJ...854L..21Z}) or the gravitational energy of an external object (e.g. asteroids hitting neutron stars; 
\citealt{2015ApJ...809...24G, 2016ApJ...829...27D})
have also been discussed in the literature. 
It is possible that not all FRB sources repeat \citep{2018ApJ...854L..12P,2019MNRAS.484.5500C}. 
If this is the case, there might be FRBs produced from catastrophic events,
such as compact star mergers
\citep{2012ApJ...755...80P,2013PASJ...65L..12T, 2013ApJ...776L..39K,2016ApJ...827L..31Z, 2016ApJ...826...82L,2016ApJ...822L...7W,2019ApJ...873L...9Z,2019ApJ...873L..13D} and
collapse of supramassive neutron stars to black holes
\citep{2014A&A...562A.137F, 2014ApJ...780L..21Z}.

The extragalactic origin of FRBs is confirmed by 
the precise localization of the first repeating FRB 121102
\citep{2014ApJ...790..101S,2016Natur.531..202S,2017ApJ...834L...8M} 
and the identification of its host galaxy
\citep{2017Natur.541...58C, 2017ApJ...834L...7T, 2017ApJ...843L...8B, 2017ApJ...844...95K}.
The host galaxy of FRB 121102 is an irregular, 
low-metallicity dwarf galaxy.
FRB 121102 resides in the bright star-forming region in the galaxy.
The properties of the host and the sub-galactic localization of the source 
is similar to those of Long Gamma Ray Bursts (LGRBs) and SuperLuminous Supernova (SLSNe), some of which have been suggested to leave behind rapidly spinning magnetars. As a result, 
young magnetars born from massive star core collapse events that produced LGRBs or SLSNe are regarded as the leading candidates to power FRBs, and it has been expected that the host galaxy of FRB 121102 should be typical for FRB sources \citep{2017ApJ...843L...8B, 2017ApJ...843...84N}. 

The search for host galaxies of other FRBs have been carried out.
FRB 150418 was proposed to be associated with a fading radio transient, 
which is located in an elliptical galaxy
\citep{2016Natur.530..453K}. 
However, the association is not secure
since the radio counterpart is a radio persistent source 
with significant variability
\citep{2016ApJ...821L..22W, 2016ApJ...824L...3A, 2016arXiv160304825L, 2016ApJ...824L...9V, 2017MNRAS.465.2143J}.
\cite{2018ApJ...867L..10M} searched for the host galaxy of FRB 171020 with a small DM (which means it is nearby) and found a host candidate ESO 601-G036. 
It is a low-metallicity Sc galaxy at redshift $z=0.00867$, which is similar to that of FRB 121102. However, the chance coincidence probability is quite large, 
and the allowable host DM of FRB 171020 is in the lower end of FRB 121102.

So it is unclear whether FRBs in general (both repeating and non-repeating ones) have host galaxies and sub-galactic environments similar to those of FRB 121102\footnote{During the review process of this paper, the host galaxies of two more FRBs, FRB 180924 \citep{2019arXiv190611476B} and FRB 190523 \citep{2019arXiv190701542R} are reported, which are different from the host of FRB 121102. This is consistent with the conclusion of our paper.}. 
We intend to investigate this problem by searching for host galaxy candidates of nearby FRBs (those with small DMs) in this paper. 
We define our nearby FRB sample in Section 2.
To prepare for the host DM estimation of the candidates,
we propose a galaxy-group-based method to estimate DM$_{\rm IGM}$ in Section 3.
We then search for the nearby FRB host candidates, 
and compare them with the host of FRB121102 in Section 4. 
We also estimate the host DM values of 
LGRB and SLSNe host galaxies, 
and compare them with those of our host candidates as well as FRB 121102
in Section 5.
We draw the conclusion that the FRB 121102 host is atypical and rare. The results are summarized in Section 6 with some discussion.
Following cosmological parameters have been adopted: 
$H_0=72.4$\,km\,s$^{-1}$\,Mpc$^{-1}$, $\Omega_{\rm DM}=0.206$, 
$\Omega_{\Lambda}=0.751$, and $\Omega_{\rm b}=0.043$
\citep{2009ApJS..180..306D}.

\section{Sample Selection}

\begin{table*}[!htb]
\begin{threeparttable}
\begin{center}
\caption{Parameters for FRBs with $\rm DM_{exc} < 100 pc\ cm^{-3}$}
\label{tb1}
\begin{tabular}{lllll|lll|llllllll}
\hline\hline
name & telescope & RA$^*$ & DEC &  DM & \multicolumn{3}{c}{NE2001} & \multicolumn{3}{|c}{YMW16} & ref \\
 & & & & pc cm$^{-3}$ & $\rm DM_{MW}$ & $\rm DM_{halo}$ & $\rm DM_{exc}$ & $\rm DM_{MW}$ & $\rm DM_{halo}$ & $\rm DM_{exc}$ & \\
\hline
180729.J1316+55 & CHIME & 13:16(28.0) & +55:32(8.0) & 109.6 & 31 & 30 & 48.6 & 22.75  & 30 & 56.9  & 1 \\
171020$^{**}$ & ASKAP & 22:15(70.5) & -19:40(62.6) & 114.1 & 38 & 30 & 46.1 & 24.71  & 30 & 59.4 &  2 \\
171213$^{**}$ & ASKAP & 03:39(47.0) & -10:56(31.3) & 158.6 & 36 & 30 & 92.6 & 40.69  & 30 & 87.9 &  2 \\
180810.J1159+83 & CHIME & 11:59(172.8) & +83:07(24.9) & 169.1 & 47 & 30 & 92.1 &  39.58  & 30 & 99.6  & 1 \\
180814.J0422+73 & CHIME & 04:22:22(4.0) & +73:40(10.0) & 189.4 & 87 & 30 & 72.4 & 108.07  & 30 & 51.3  & 3 \\

\hline
\end{tabular}
\begin{tablenotes}
\small
\item $^*$ Positional uncertainties are 99\% confidence limits and in unit of arcminute.
The positional uncertainties of 180729.J1316+55 are given three times in \cite{2019Natur.566..230C}, as (21$'$, 8$'$), (28$'$, 12$'$), and (28$'$, 8$'$).
We use (28$'$, 8$'$) here.
\item $^{**}$Positional information are obtained from https://data.csiro.au/collections/\#collection/CIcsiro:34437v3
\item reference: (1) \cite{2019Natur.566..230C};
(2) \cite{2018Natur.562..386S};
(3) \cite{2019Natur.566..235C}
\end{tablenotes}
\end{center}
\end{threeparttable}
\end{table*}

We use the DM values of FRBs to select nearby FRBs.
We decompose the total observed DM into four terms:
$$\rm DM_{tot}=DM_{\rm MW}+DM_{\rm halo}+DM_{\rm IGM}+DM_{\rm host},$$
where $\rm DM_{\rm MW}$ is the contribution from the Milky Way disk,
which is estimated using the NE2001 \citep{2002astro.ph..7156C}
or YMW16 \citep{2017ApJ...835...29Y} models constructed with the observed pulsar DM data; $\rm DM_{halo}$ is the contribution from Milky Way halo, which is
estimated to be 30\,pc\,cm$^{-3}$ in \cite{2015MNRAS.451.4277D}
or $50-80$\,pc\,cm$^{-3}$ in \cite{2019MNRAS.485..648P} from simulations -- 
to be conservative, we used 30 pc cm$^{-3}$ for our estimation; and
$\rm DM_{\rm IGM}$ and $\rm DM_{\rm host}$ are the contributions from the intergalactic medium (IGM) and from the host galaxy, respectively. The latter
also includes the contribution from the FRB local environment.
We would like to use $\rm DM_{IGM}$ to constrain the distance,
and investigate $\rm DM_{host}$ in this paper.
We select the FRBs with the excess DM,
\begin{eqnarray}
\rm DM_{exc} & = & \rm DM_{tot}-DM_{MW}-DM_{halo} \\
             & = & \rm DM_{IGM}+DM_{host} < 100\ pc\ cm^{-3},
\end{eqnarray}
from the FRBCAT catalog\footnote{www.frbcat.org}.
There are 5 in total, whose basic information is listed
in Table \ref{tb1}. The second repeating FRB discovered by \cite{2019Natur.566..235C}, FRB 180814.J0422+73, is also on the list.
We convert their positional uncertainties to 99\% confidence level
based on the Gaussian distribution, which are presented in units of arcminutes in Table \ref{tb1}.

\section{IGM DM}

\begin{figure*}[!htb]
\centering
\includegraphics[width=0.90\columnwidth]{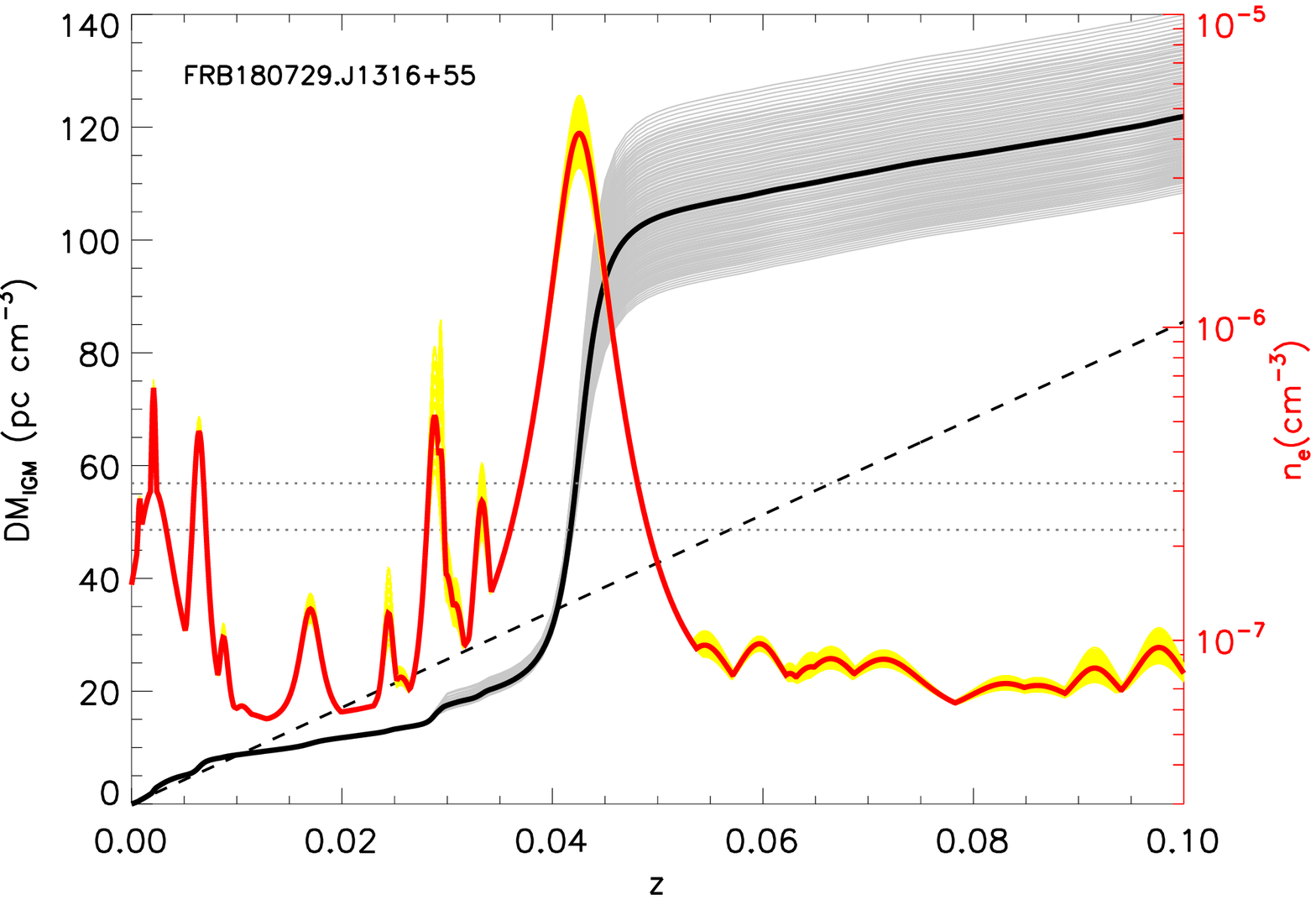}
\includegraphics[width=0.90\columnwidth]{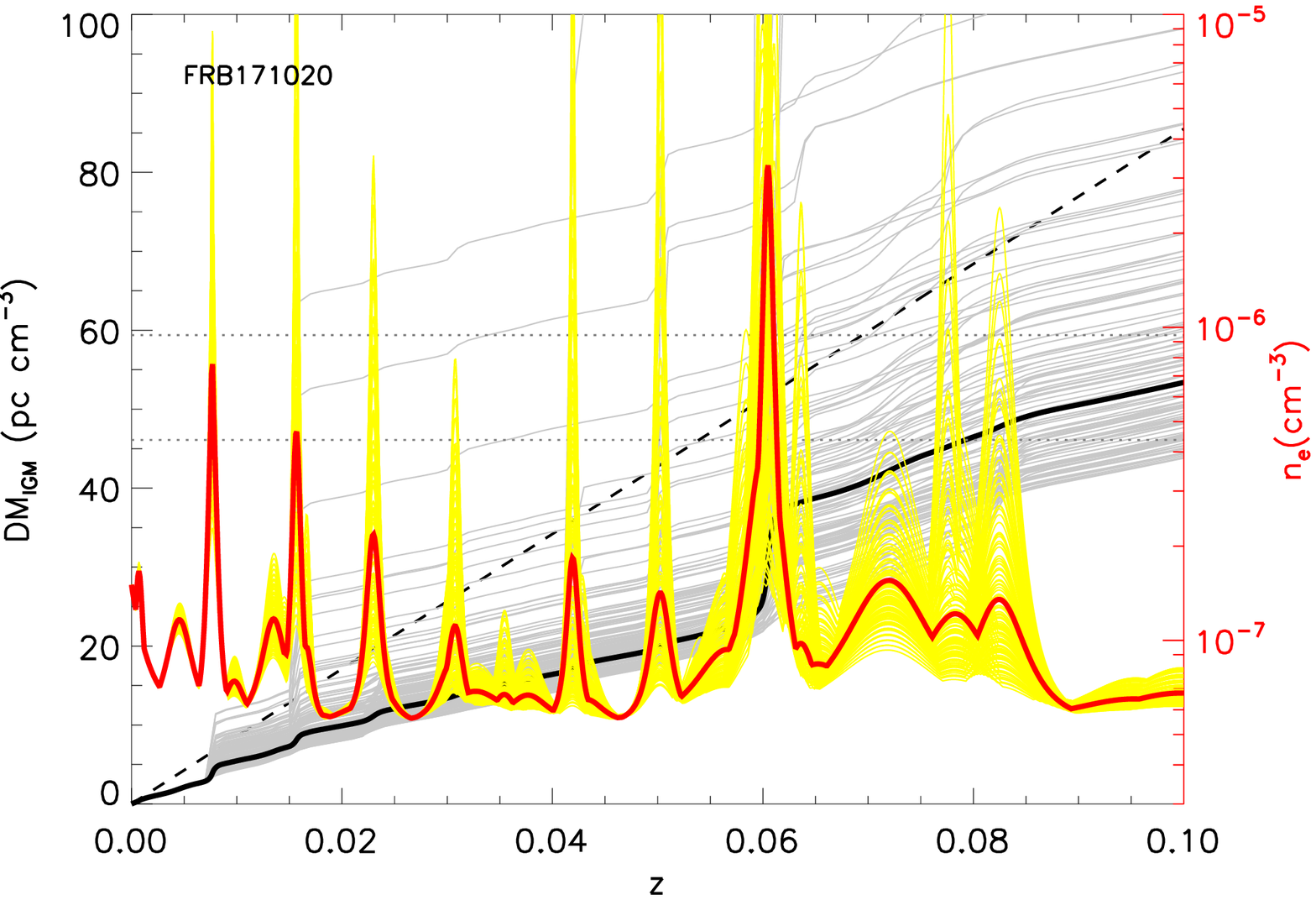}
\includegraphics[width=0.90\columnwidth]{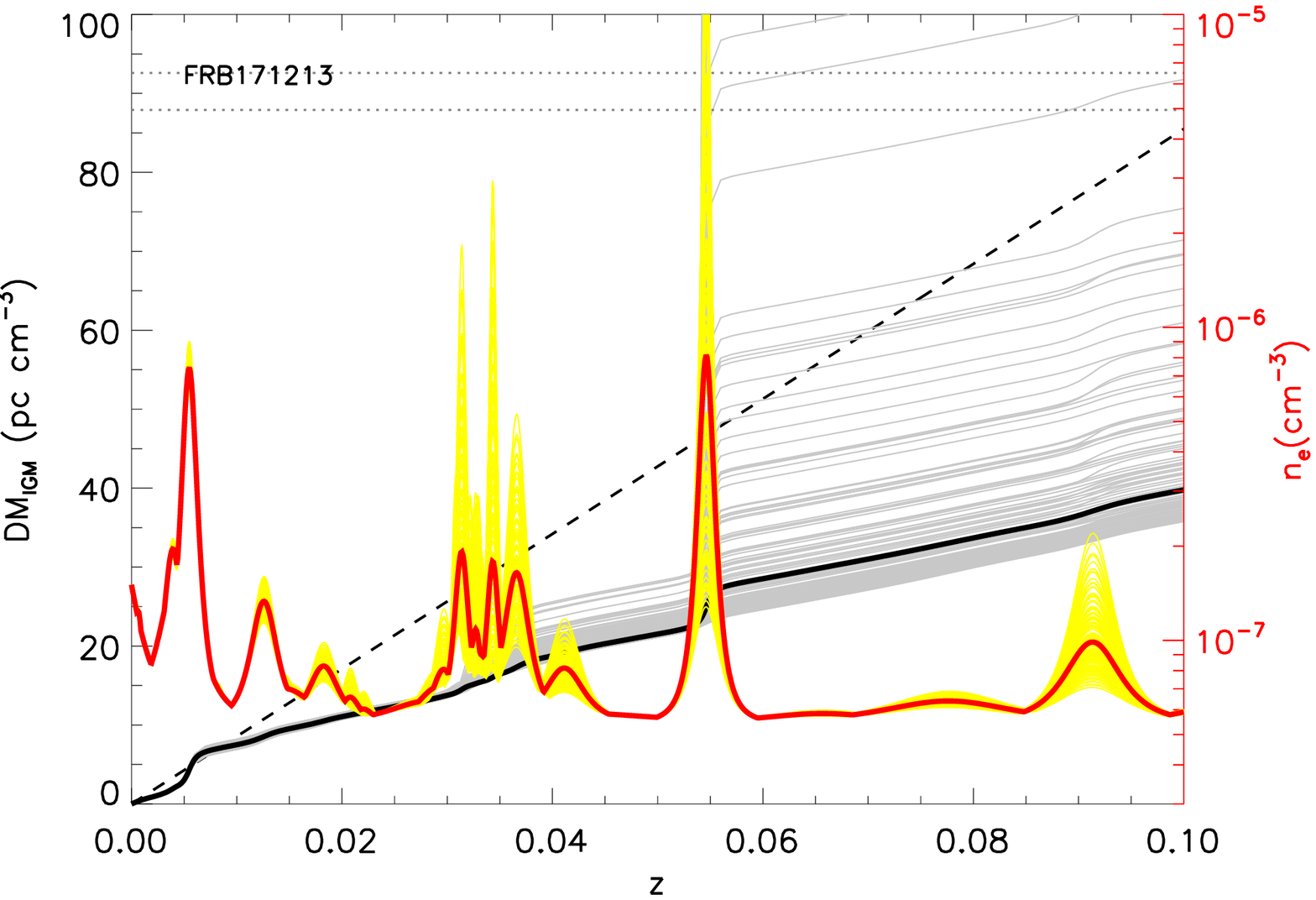}
\includegraphics[width=0.90\columnwidth]{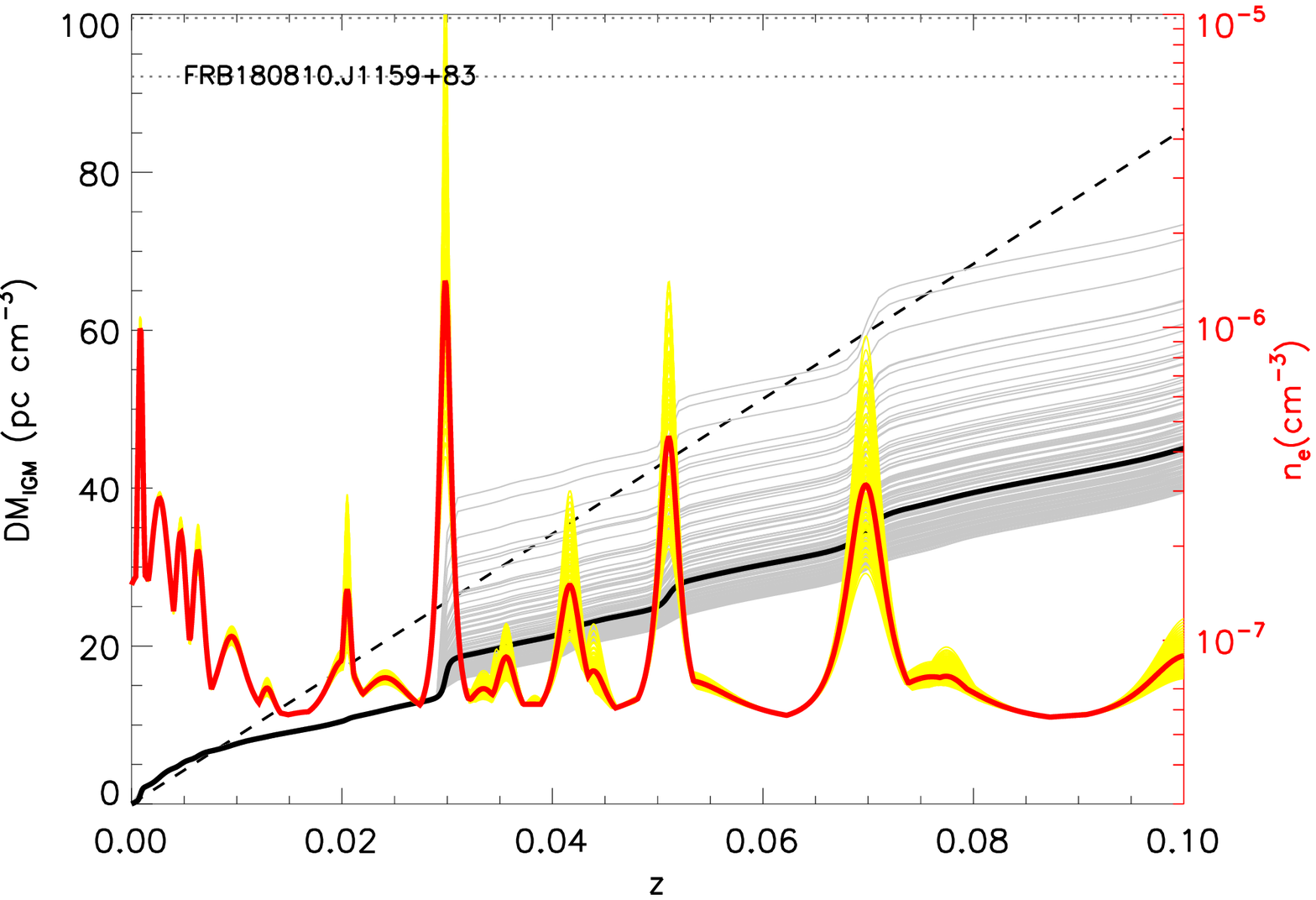}
\includegraphics[width=0.90\columnwidth]{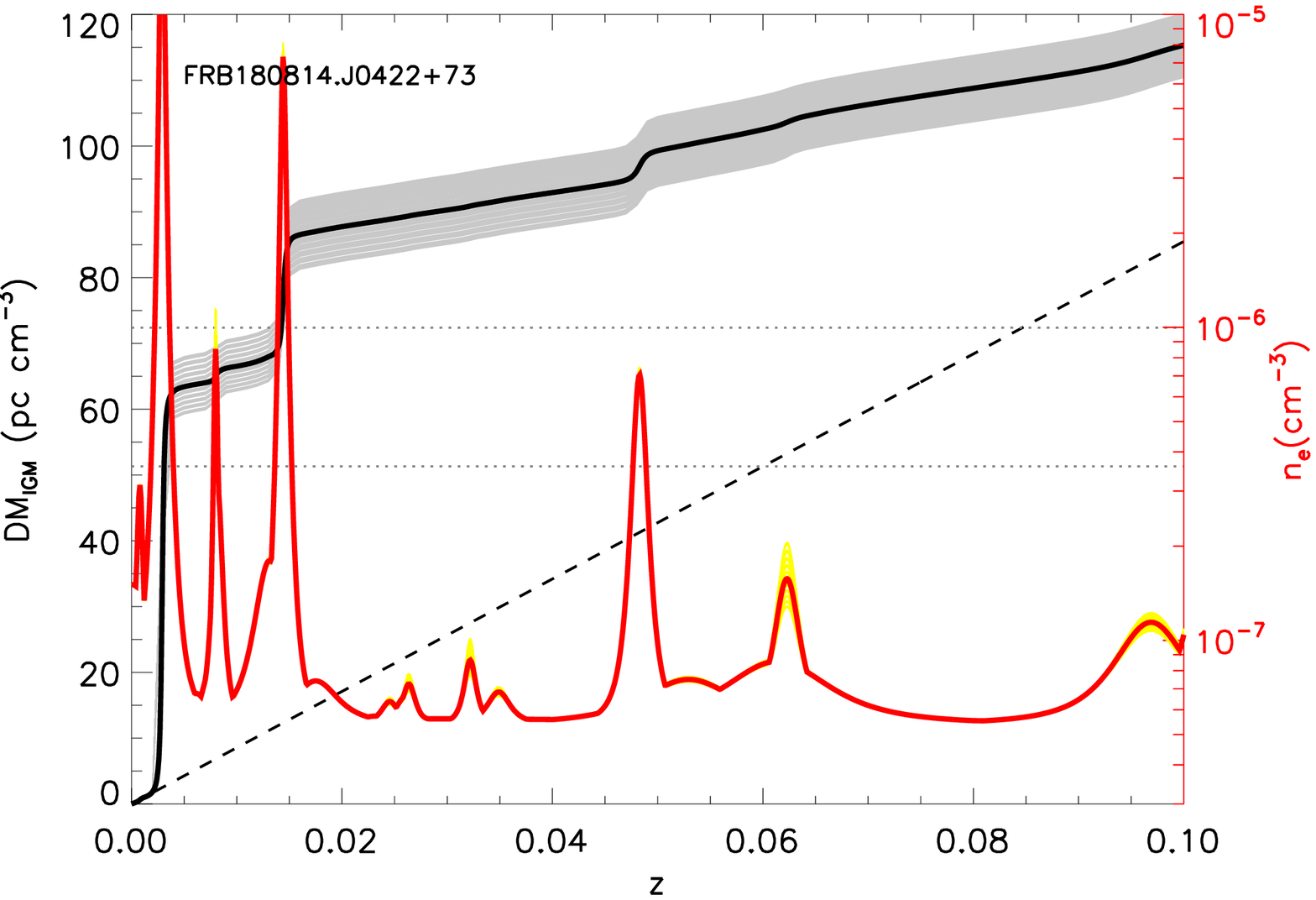}
\includegraphics[width=0.90\columnwidth]{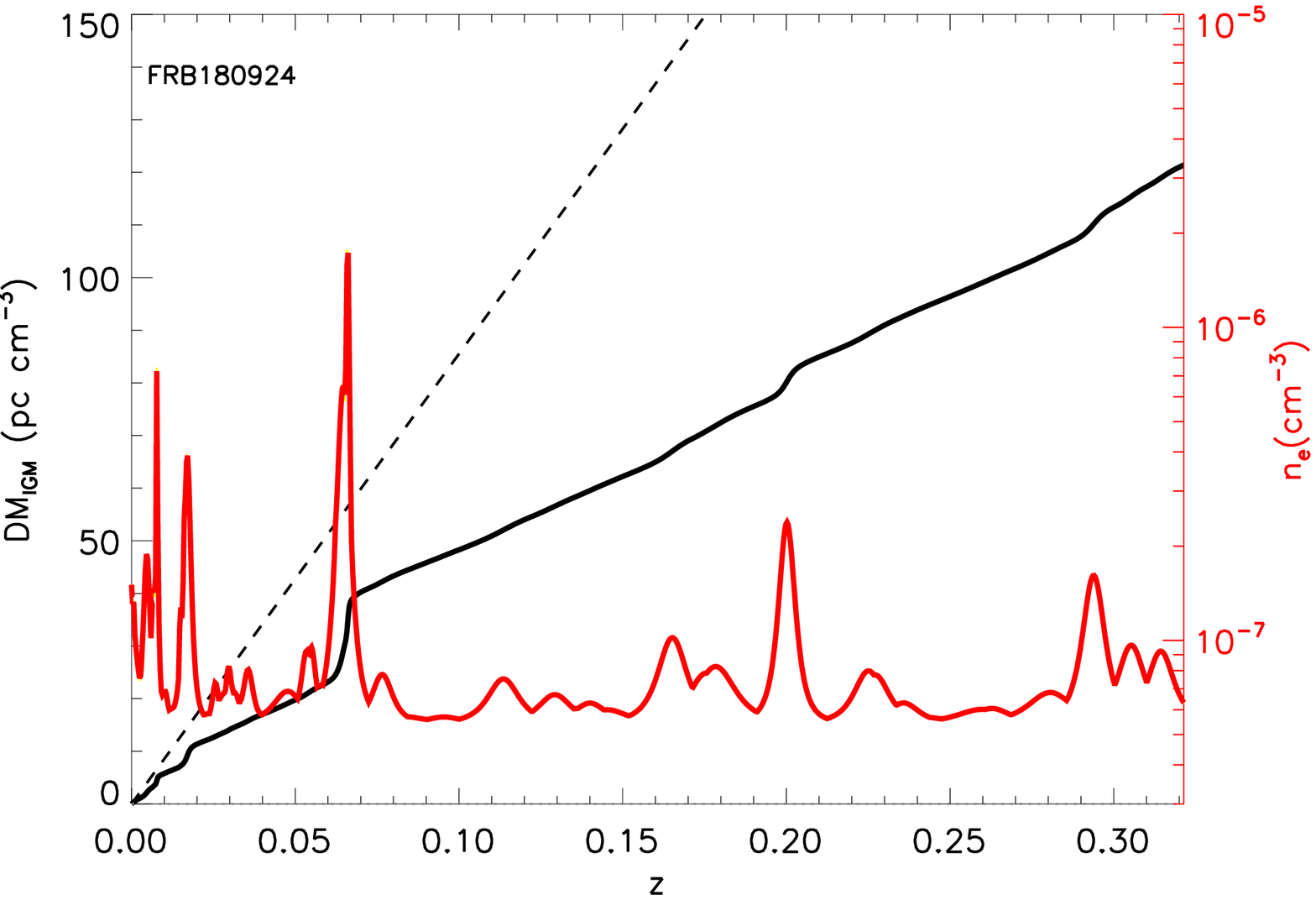}
\includegraphics[width=0.90\columnwidth]{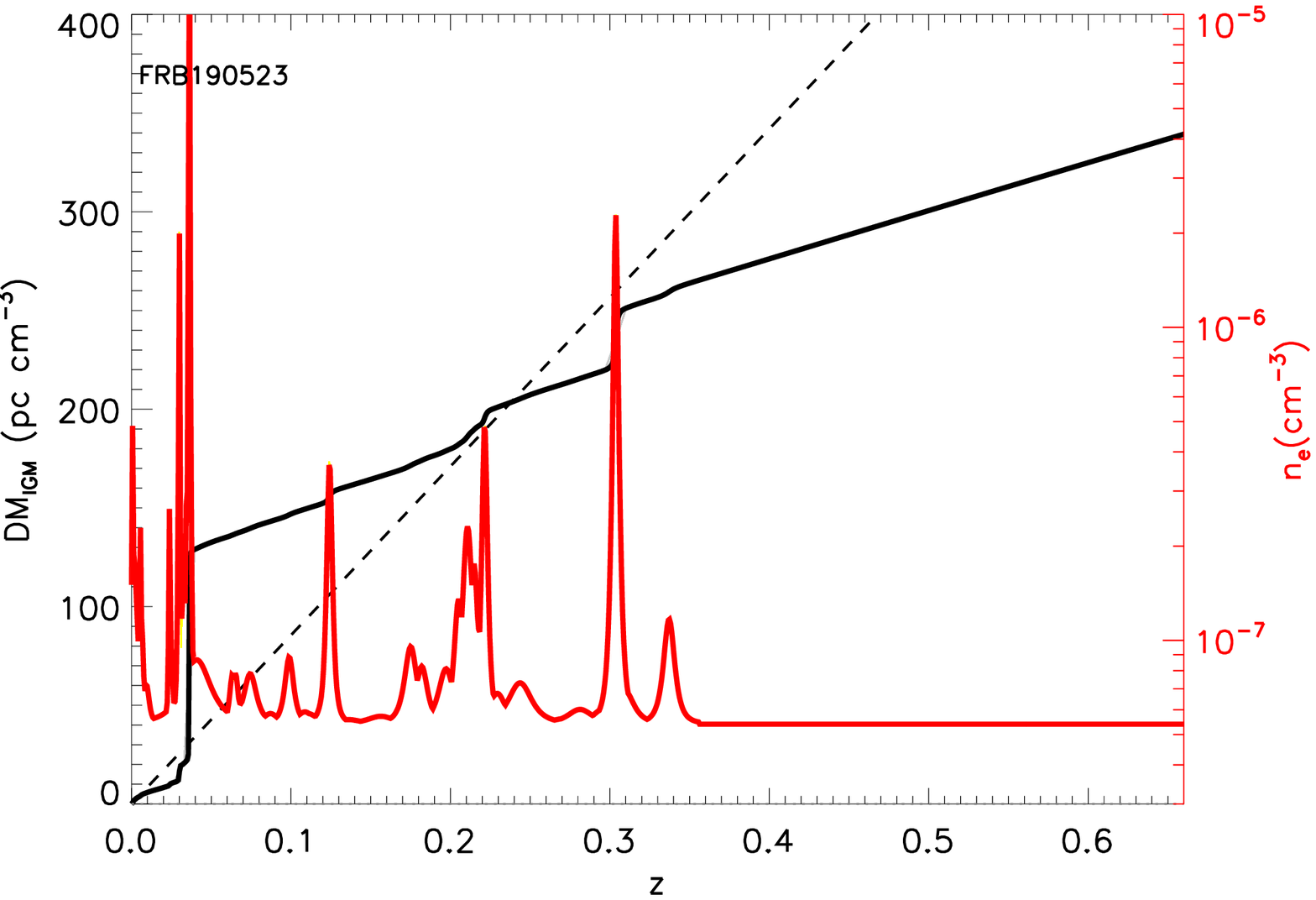}
\includegraphics[width=0.90\columnwidth]{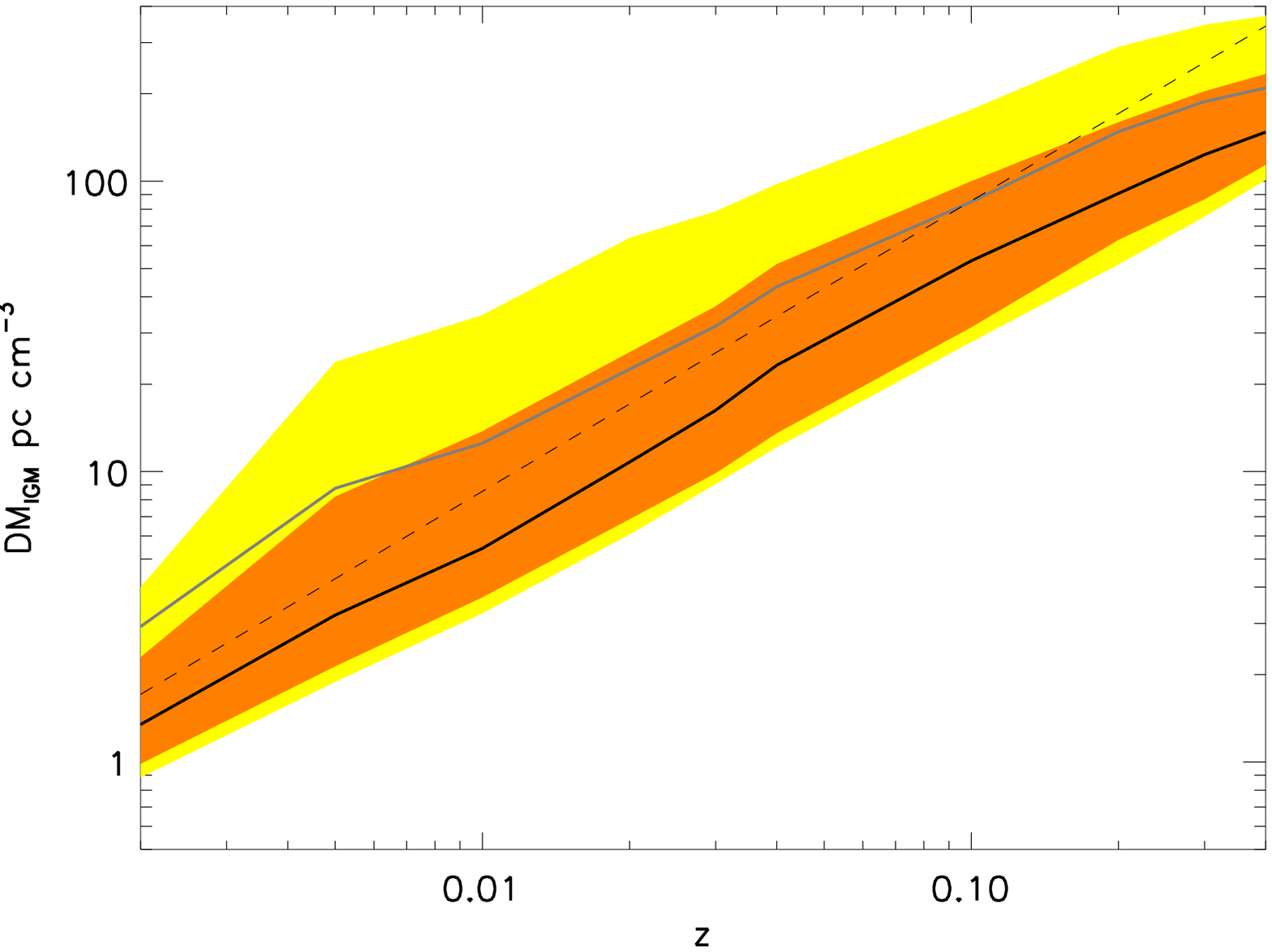}
\caption{
{The DM$_{\rm IGM}$ (black and grey lines) and electron density (red and yellow lines) as a function of redshift estimated with the galaxy group catalogs.
The thick black lines and thick red lines are for the positional center of each FRB. 
Yellow and grey lines show the values for difference positions within the positional uncertainty for each FRB. 
}
The empirical ${\rm DM_{IGM}}=855z$ pc cm$^{-3}$ relation is
presented as the dashed line for comparison.
The range of $\rm DM_{exc}$ is shown as the horizontal dotted lines.
Lowest right panel: 
The distribution of DM$_{\rm IGM}$ as a function of $z$. 
The black thick line and the grey thick line are the median and mean values. 
Orange and yellow areas are 68\% and 90\% regions, respectively. 
Again, The empirical $\rm DM_{IGM}=855z$ pc cm$^{-3}$ relation is
presented as the dashed line. 
}
\label{fig:dmz}
\end{figure*}

In order to investigate the allowable redshift range,
and estimate the host DM,  
a relation between redshift and $\rm DM_{IGM}$, i.e. ${\rm DM_{IGM}}=fz$, is usually applied
\citep{2003ApJ...598L..79I, 2004MNRAS.348..999I,2014ApJ...783L..35D, 2018ApJ...867L..21Z}, 
with $f$ in the range of $\sim 850 - 1200$\,pc\,cm$^{-3}$.
However, the relation between $\rm DM_{IGM}$ and redshift suffers from large uncertainties. 
Cosmological simulations reveal that the line-of-sight fluctuations dominate 
the DM$_{\rm IGM}$ uncertainties. 
The difference resulting from different lines of sight can be substantial 
\citep{2014ApJ...780L..33M, 2019MNRAS.484.1637J, 2019arXiv190307630P}.

In order to eliminate the line-of-sight uncertainty, 
here we propose to directly use the observed galaxy group information to estimate the cosmic density field, and hence $\rm DM_{IGM}$, along the lines of sight of FRBs in our sample.

With the observed galaxy groups, \cite{2009MNRAS.394..398W,2016ApJ...831..164W}
developed a halo-domain method to reconstruct the cosmic density field.
However, \cite{2016ApJ...831..164W} only covered the SDSS DR7 region, 
which contains only one of our object, FRB 180729.J1316+55.
We therefore use a nearly all-sky galaxy-group catalog in \cite{2017MNRAS.470.2982L}. To the first-order estimate, we adopt
the empirical Navarro-Frenk-White (NFW) dark matter density profile
\citep{1997ApJ...490..493N} to reconstruct the cosmic density field. 

There are four galaxy group catalogs in \cite{2017MNRAS.470.2982L}, 
which are produced with the galaxy catalogs from 2MASS redshift survey (2MRS), 
6dF Galaxy Survey (6dFGS), Sloan Digital Sky Survey (SDSS), and 
2dF Galaxy Redshift Survey (2dFGRS). 
Among them, 2MRS has the 91\% sky coverage, nearly all except the galactic plane.
Most of our FRBs are only covered by 2MRS so
we use the 2MRS catalog here.
However, redshift is not a good indicator of distance 
for nearby galaxies. 
We thus update the distance of the galaxies with the nearby galaxy catalog
of \cite{2013AJ....145..101K}, and then propagate the updated distances
to the corresponding groups.
Note that 2MRS is not complete for $z>0.033$
\citep{2015AJ....149..171T}.
The DM$_{\rm IGM}$ with $z>0.033$ is only a lower limit of DM.

For each of the galaxy groups, \cite{2017MNRAS.470.2982L} estimated 
their dark matter halo masses, log $M_{\rm h}$ ($M_{\odot}$ h$^{-1}$). 
For each dark matter halo,
we estimate the dark matter density profile as
\begin{equation}
\rho(r)=\frac{\rho_0}{(r/R_{\rm s})(1+r/R_{\rm s})^2},
\end{equation}
where $r$ is the distance from the center. 
$R_{\rm s}=r_{200}/c_{200}$ is a scale radius.
$r_{200}$ is the radius where the average density of the halo
is 200 times of cosmic critical density, $\rho_{\rm c}=3H^2/8\pi G$;
$c_{200}$ is the concentration of the halo, depending the halo mass and redshift,
and we use $\log c_{200}=0.830-0.098\ {\rm log}M_{\rm h}$ from 
\cite{2008MNRAS.391.1940M}.
The normalization is $\rho_0=M_{\rm h}/\{4\pi R_{s}^3[{\rm ln}(1+c)-c/(1+c)]\}$. 
However, the observational group catalog is flux limited.
\cite{2009MNRAS.394..398W, 2016ApJ...831..164W} revealed that
haloes with masses smaller than $10^{12}\ \rm M_{\odot}\ h^{-1}$
continues smoothly to the background density of 
0.2 times the mean mass density ($\rho_{\rm m} = \Omega_{\rm m} \rho_{\rm c}=2.45 \times 10^{-30}\rm\ g\ cm^{-3}$ for $h=0.724$) of the Universe,
where $\Omega_{\rm m}$ is the normalized mass density.
We thus limit our galaxy groups to those with halo masses larger than $10^{12}\ \rm M_{\odot}\ h^{-1}$,
and consider $\rho = 0.2\rho_{\rm m}$ as the background density 
in the intergalactic space in addition to the NFW density profile for the groups.

We convert the dark matter mass density $\rho$ to baryon mass density by the ratio between $\Omega_{\rm b}$ and $\Omega_{\rm dm}$, 
the normalized baryon and dark matter mass densities.
If the baryon in the IGM traces dark matter and is composed of totally ionized hydrogen and helium, then
the free electron number density $n_{\rm e}$ can be related to the dark matter density $\rho$ by
\begin{eqnarray}
n_{\rm e} & = & \rho\frac{\Omega_{\rm b}}{\Omega_{\rm dm}}\left (\frac{Y_{\rm H}}{m_{\rm H}}+2\frac{Y_{\rm He}}{m_{\rm He}} \right) 
 =  \rho\frac{\Omega_{\rm b}}{\Omega_{\rm dm}}\frac{0.875}{m_{\rm H}} \\
& = & 2.73 \times 10^{-7} {\rm cm}^{-3} \frac{\rho}{\rho_{\rm m}},
\end{eqnarray}
where $Y_{\rm H}=0.75$ and $Y_{\rm He}=0.25$ are the mass fractions of hydrogen and helium, and $m_{\rm H}$ and $m_{\rm He}$ are the masses of their atoms.

The electron density as a function of redshift for our nearby FRBs
is presented in Figure \ref{fig:dmz}. 
The red curve represents the electron density $n_{\rm e}$ as a function of distance (redshift) at the center of the positional region for each FRB.
{The yellow lines represent other $11 \times 11$ lines of sight within the positional uncertainties of each FRB.
The black curve shows $\rm DM_{IGM}$ as a function of redshift at the center of
the position uncertainty. The grey lines are again for other lines of sight within the positional uncertainties. }
The ${\rm DM_{IGM}}=855z$ pc cm$^{-3}$ relation \citep{2018ApJ...867L..21Z} is also plotted as the dashed line for comparison. It can be seen that for individual FRB sources, ${\rm DM_{IGM}}$ can be much deviated from the average value. The line of sight of FRB 180729.J1316+55 goes through a massive galaxy group around redshift 0.04. Its $\rm DM_{IGM}$ reaches much more than predicted by the empirical ${\rm DM_{IGM}}=855z$ pc cm$^{-3}$ relation at around $z=0.05$. The largest redshift of its host galaxy is around 0.05.
The center lines of sight of FRB 171020, FRB 171213 and FRB 180810.J1159+83 
only go through the edge of their respective galaxy groups.
Therefore, there are only small peaks in their electron density curves, 
and their ${\rm DM_{IGM}}$ values are smaller than the dashed line.
However, since their positional uncertainties are large, 
it is still possible that their lines of sight indeed pass
through galaxy groups or even the center of the groups.
In such cases, their ${\rm DM_{IGM}}$ values are boosted a lot, 
even higher than 100 pc cm$^{-3}$.
The line of sight of FRB180814.J0422+73, the second repeating FRB, 
goes through many galaxy groups within $z=0.02$.
Its ${\rm DM_{IGM}}$ is larger than the value from the ${\rm DM_{IGM}}=855z$ relation even if the 2MRS catalog is incomplete.
Its ${\rm DM_{IGM}}$ reaches $\rm DM_{exc}$ around $z=0.01$, indicating that
its host is likely extremely nearby.

{For comparison, We have also examined FRB 121102. 
However, FRB 121102 is too close to the Galactic Plane, with a galactic latitude $-0.2$ degree. 
This region is avoided by most galaxy group catalogs.
So, we are unable to constrain its ${\rm DM_{IGM}}$.
}

To compare with other cosmological results,
we calculate $\rm DM_{IGM}$ for different redshifts and all sky, with 360 bins in RA, and 180 bins in DEC. 
The distribution of the $\rm DM_{IGM}$ as a function of redshift $z$ is plotted in the lowest right panel of Fig. \ref{fig:dmz}.
The black thick curve indicates the median value for each redshift, 
and the grey curve presents its mean value. 
The orange and yellow regions show the 68\% and 90\% confidence levels, respectively.
The black dashed curve is again ${\rm DM_{IGM}}=855z$ relation. 
It turns out that the median and mean values bracket 
the ${\rm DM_{IGM}}=855z$ relation with $z<0.033$, 
and follows nearly the same shape. 
It indicates that our result is generally consistent with 
previous rough estimation by \cite{2018ApJ...867L..21Z}, 
and our 2MRS galaxy group sample is generally complete at $z<0.033$.
However, our results flatten when reaching redshift 0.04 
due to the incompleteness of 2MRS at higher redshifts.
Thus, our estimation should be considered as the lower limit for 
$z > 0.033$.

Even without knowing the true redshift, our analysis gives a relation between DM$_{\rm IGM}$ and $z$ for individual FRBs with certain uncertainties.
With such a preparation, we can then estimate the values of the host DM, i.e. $\rm DM_{host}=DM_{exc}-DM_{IGM}$, of each FRB for different redshifts.
For $z>0.033$, our derived $\rm DM_{host}$ can be regarded as the upper limits.
These derived values can be then compared with that of FRB121102 (Section \ref{candi} next).

\section{Host galaxy candidates} \label{candi}

\begin{table*}[!htb]
\begin{threeparttable}
\begin{center}
\caption{Parameters for host galaxy candidates with redshift measurements}
\label{tb:candi}
\begin{tabular}{llllllllllllllll}
\hline\hline
name & RA & DEC & redshift & $\rm DM_{IGM}$ & $\rm DM_{host, NE2001}$ & $\rm DM_{host, ymw16}$ & m$_{\rm g}$ & M$_{\rm B}$ \\\hline
\hline FRB180729.J1316+55 & 7/59/695 \\\hline
SDSSJ131613.66+553741.5 & 199.05799 & 55.63030 & 0.0270 & 13.9 & 34.7 & 43.0 & 16.7 & -18.7 \\
2MASSJ13170558+5529488 & 199.27356 & 55.49705 & 0.0394 & 28.4 & 20.2 & 28.5 & 17.0 & -19.1 \\
SDSS J131436.14+553530.2 & 198.65062 & 55.59173 & 0.0810 & 115.3 & --- & --- & 17.8 & -19.9 \\
SDSSJ131440.13+552402.8 & 198.66723 & 55.40073 & 0.0827 & 115.8 & --- & --- & 17.7 & -20.1 \\
2MASSJ13144317+5535576 & 198.67964 & 55.59920 & 0.1138 & 131.6 & --- & --- & 18.2 & -20.3 \\
SDSS J131539.49+552817.0 & 198.91455 & 55.47140 & 0.1193 & 136.4 & --- & --- & 18.2 & -20.4 \\
SDSS J131720.0+553021.2 & 199.33329 & 55.50588 & 0.1247 & 141.0 & --- & --- & 22.0 & -16.7 \\
\hline FRB171020 & 12/31/4974 \\\hline
ESO 601- G 036 & 333.85350 & -19.58519 & 0.0087 & 5.1 & 41.0 & 54.3 & 15.2 & -17.7 \\
2MASSJ22172928-1954557 & 334.37205 & -19.91542 & 0.0514 & 20.6 & 25.5 & 38.8 & 16.5 & -20.3 \\
2MASSJ22131992-2002022 & 333.33304 & -20.03384 & 0.0619$^*$ & 37.3 & 8.8 & 22.1 & 16.5 & -20.6 \\
2MASSJ22171676-1901556 & 334.31987 & -19.03206 & 0.0628 & 37.9 & 8.2 & 21.5 & 16.2 & -21.0 \\
2MASSJ22165509-1934325 & 334.22969 & -19.57576 & 0.0632$^*$ & 38.0 & 8.1 & 21.4 & 17.0 & -20.2 \\
2MASSJ22150112-1925373 & 333.75481 & -19.42699 & 0.0666 & 39.3 & 6.8 & 20.1 & 16.3 & -21.1 \\
2MASSJ22161241-1909585 & 334.05162 & -19.16632 & 0.0832$^*$ & 48.2 & --- & 11.2 & 17.3 & -20.5 \\
2MASSJ22160049-1900395 & 334.00186 & -19.01089 & 0.0923$^*$ & 51.2 & --- & 8.2 & 17.2 & -20.9 \\
2MASSJ22164473-1903516 & 334.18648 & -19.06445 & 0.0925$^*$ & 51.2 & --- & 8.2 & 17.0 & -21.1 \\
2MASSJ22132225-1947211 & 333.34281 & -19.78928 & 0.1030$^*$ & 56.0 & --- & 3.4 & 17.5 & -20.8 \\
2MASSJ22153780-2033247 & 333.90750 & -20.55684 & 0.1074$^*$ & 59.8 & --- & --- & 17.3 & -21.1 \\
2MASSJ22145283-2008131 & 333.72019 & -20.13693 & 0.1378$^*$ & 85.7 & --- & --- & 18.0 & -20.9 \\
\hline FRB171213 & 5/8/1963 \\\hline
2MASSJ03412673-1031406 & 55.36138 & -10.52779 & 0.1059$^*$ & 44.9 & 47.7 & 43.0 & 17.4 & -21.1 \\
2MASSJ03383757-1109423 & 54.65652 & -11.16177 & 0.1368$^*$ & 71.2 & 21.4 & 16.7 & 17.1 & -22.0 \\
2MASSJ03414775-1026428 & 55.44890 & -10.44525 & 0.1400$^*$ & 73.9 & 18.7 & 14.0 & 17.9 & -21.2 \\
2MASSJ03385211-1058563 & 54.71704 & -10.98223 & 0.1406$^*$ & 74.5 & 18.1 & 13.4 & 18.4 & -20.8 \\
2MASSJ03382824-1104255 & 54.61758 & -11.07368 & 0.1409$^*$ & 74.7 & 17.9 & 13.2 & 18.1 & -21.0 \\
\hline FRB180810.J1159+83 & 3/3/1066 \\\hline
2MASSJ11552291+8246314 & 178.84550 & 82.77529 & 0.0438$^*$ & 23.0 & 69.1 & 76.5 & 16.6 & -20.4 \\
2MASSJ12045319+8322007 & 181.22218 & 83.36675 & 0.0816$^*$ & 39.9 & 52.2 & 59.7 & 17.7 & -20.6 \\
2MASSJ11595630+8301545 & 179.98360 & 83.03170 & 0.1203$^*$ & 62.4 & 29.7 & 37.1 & 18.3 & -20.9 \\
\hline FRB180814.J0422+73 & 1/1/50 \\\hline
2MASSJ04222144+7347101 & 65.58900 & 73.78612 & 0.0781$^*$ & 108.3 & --- & --- & 17.5 & -20.6 \\
\hline
\end{tabular}
\begin{tablenotes}
\small
\item $^*$ Photometric redshift from 2MPZ.
\end{tablenotes}
\end{center}
\end{threeparttable}
\end{table*}

\begin{figure*}[!htb]
\centering
\includegraphics[width=1.5\columnwidth]{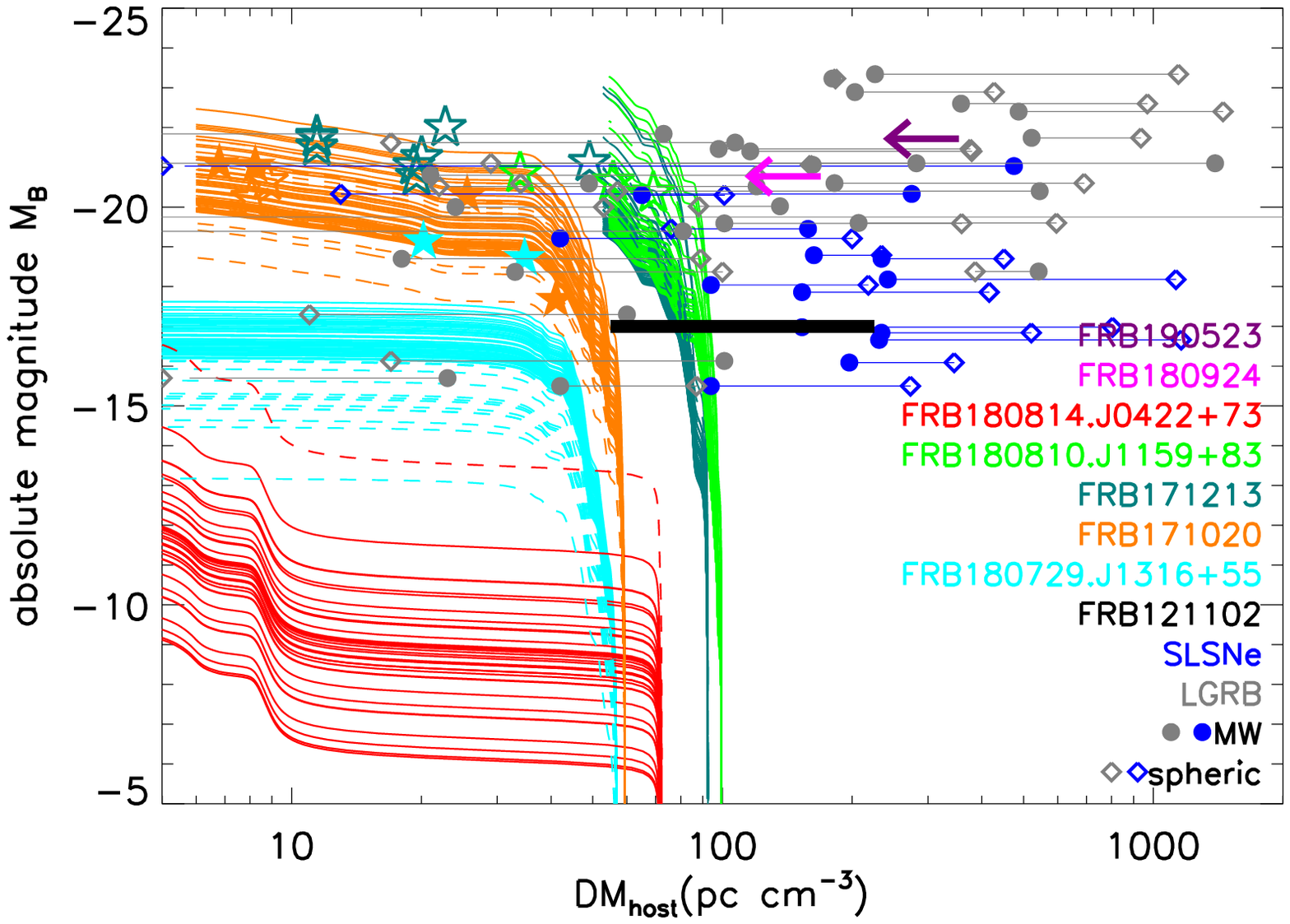}
\caption{
B band absolute magnitudes $M_{\rm B}$ versus $\rm DM_{host}$.
Candidates for different FRBs are plotted with different colors.
Filled stars indicate candidates with spectroscopic redshifts.
For FRB candidates without redshift information, 
we assume different redshifts in $0-0.1$ to calculate $M_{\rm B}$ and $\rm DM_{host}$ of them, and plot as solid lines. 
Candidates with photometric redshifts are presented as open stars, 
and they are also plotted as dashed lines for different assumed redshifts.
For comparison, FRB 121102 is plotted as a thick black solid line.
{FRB180924 and FRB190523 are presented as magenta and purple arrows. Their $\rm DM_{host}$ are estimated based on the estimation of $\rm DM_{IGM}$.}
LGRB and SLSNe host galaxies are denoted as grey and blue symbols.
Filled dots are for the MW template, 
and diamonds are for the spherical electron density profile. 
}
\label{fig:DMhost_aMag}
\end{figure*}

We search for host galaxy candidates using RA, DEC of each FRB and its 99\% errors.
For FRB 171020 and FRB 171213, the localization probability images provided by
\cite{2018Natur.562..386S} are employed.
Since our FRBs are expected to be nearby,
we first explore the Galaxy List for the Advanced Detector Era (GLADE) catalog
\citep{2018MNRAS.479.2374D}.
It is a nearby galaxy catalog
aiming at providing host galaxy candidates to Gravitational Wave events.
It combines the galaxies in Gravitaional Wave Galaxy Catalog (GWGC) \citep{2011CQGra..28h5016W},
2MASS Photometric redshift catalog (2MPZ) \citep{2014ApJS..210....9B}\footnote{http://ssa.roe.ac.uk/TWOMPZ.html},
2MASS extended source catalog (2MASS XSC) \citep{2006AJ....131.1163S},
HyperLEDA \citep{2014A&A...570A..13M}, and SDSS-DR12Q \citep{2017A&A...597A..79P}.
For each host candidate, we double check the redshift information 
in SDSS\footnote{http://skyserver.sdss.org/dr15/en/tools/chart/navi.aspx} and NED\footnote{http://ned.ipac.caltech.edu}.

In order to be more complete, we also explore the
extended sources in the Pan-STARRS catalog\footnote{https://outerspace.stsci.edu/display/PANSTARRS}
\citep{2016arXiv161205560C, 2016arXiv161205243F}.
Following the menu of Pan-STARRS,
we select objects in StackObjectThin database table,
exclude spurious sources by requiring ndetections $>$ 1,
and select Pan-STARRS galaxies by requiring
$\rm mag_{PSF}-mag_{kron} > 0.05$.
We find and delete duplicate objects whose coordinates are off by 1 arcsecond.
We then assign redshifts from SDSS, 2MPZ and NED to Pan-STARRS sources, allowing a coordinate offset by 3 arcseconds. 

We use RA, DEC, and their 99\% errors to select host galaxy candidates.
For FRB 171020 and FRB 171213, we use the localization probability images provided by \cite{2018Natur.562..386S} to select the galaxy candidates.
The candidates with redshifts, spectroscopic or photometric, less than 0.15
are presented in Table \ref{tb:candi}.
The numbers after the FRB names give the number of galaxies 
with redshift less than 0.15, the number of galaxies with redshifts, 
and the total number of galaxies (most of them are Pan-STARRS extended sources).
For each of the candidates, 
we estimate the expected $\rm DM_{host}=DM_{exc}-DM_{IGM}$ for 
NE2001 and YMW16, respectively, 
as $\rm DM_{host,NE2001}$ and $\rm DM_{host,ymw16}$.
For those with redshift $z>0.1$, 
the $\rm DM_{IGM}$ is estimated by requiring
$\Delta {\rm DM_{IGM}}=855 \Delta z$ for $z>0.1$, 
and following Fig. \ref{fig:dmz} for $z<0.1$.
With their $g$-band Kron magnitude (when available), or $g$-band PSF magnitudes, presented as $m_g$,
we estimate their
$B$-band absolute magnitude $M_{\rm B}$ following
$M_{\rm B}=m_{g}-A_g-5{\rm log}(\frac{D_{\rm L}}{10{\rm pc}})-2.5(\beta+2){\rm log}\left[\frac{(1+z)\lambda_0}{\lambda}\right]+2.5{\rm log}(1+z)$
\citep{2011ApJ...739....1L},
where $A_g$ is the Galactic extinction in $g$ band, $D_{\rm L}$ is luminosity distance and $z$ is the redshift, 
$\lambda$ is the observational effective wavelength, 4900 \AA\ for PanSTARRS $g$ band, $\lambda_0$ is 
the rest frame effective wavelength, 4300 \AA\ for $B$ band, and  
$\beta$ is the index of the assumed power law spectrum, $F_{\lambda} \propto \lambda^{\beta}$. Because we examine blue/green bands and the expected spectrum of FRB 121102 host may be similar to GRB hosts, we adopt $\beta=-2.3$ following \citealt{2011ApJ...739....1L}. 
Their $g$-band magnitude and absolute magnitudes $M_{\rm B}$
are also presented.

The candidates are compared with the host of FRB 121102 in a $\rm DM_{host}$ vs. absolute magnitude $M_B$ diagram (Fig. \ref{fig:DMhost_aMag}).
\cite{2017ApJ...834L...7T} estimated the $\rm DM_{host}$ of FRB121102 
to be $\rm DM_{host}=140\pm85$ pc cm$^{-3}$, 
by an empirically estimated $\rm DM_{IGM}$, with an error of 85 pc cm$^{-3}$.
\cite{2017ApJ...844...95K} gives $\rm DM_{host}=163 \pm 96\rm\ pc\ cm^{-3}$
by taking the uncertainty of MW, IGM and observation into account.
We thus use $55<\rm DM_{host}<225$ pc cm$^{-3}$ to be conservative,
and presented it as the black thick line.
Candidates for different FRBs are represented 
by different colors.
Candidates with spectroscopic redshifts are plotted as filled stars,
with the values in Table \ref{tb:candi}.
For those without redshifts, 
we estimate their $\rm DM_{host}$ and $M_{\rm B}$ by assuming redshifts $z=0-0.1$, following the same method in the last paragraph, and then plot them as solid curves. To be clear, we only plot the brightest 50 candidates without redshifts
for each FRB. There are many more galaxies fainter than what we presented.
Candidates with photometric redshifts are presented as open stars,
with dashed curves indicating different redshifts also.

\subsection{FRB 180729.J1316+55}

There are 695 extended sources within the positional region of FRB180729.J1316+55. 
Among them, 59 have spectroscopic redshifts from SDSS.
There are 7 galaxies with spectroscopic redshifts less than 0.15.
Two of them, SDSSJ131613.66+553741.5 and 2MASS13170558+5529488
have relatively large values of $\rm DM_{host}$\footnote{GLADE used the photometric redshift 0.06263 in the catalog. However, its SDSS spectroscopic redshift is 0.039.}.
 
The first one, SDSSJ131613.66+553741.5, is a faint source within a big disk galaxy 
SDSSJ131613.95+553749.5. It is likely a star forming region in the galaxy.
Its expected $\rm DM_{host}$ 
is 35 pc cm$^{-3}$ and 43 pc cm$^{-3}$ for the NE2001 and YMW16 models, respectively. 
The second one, SDSS131705.58+552948.8, is an edge-on disk galaxy 
with a significant bulge. 
{SED fitting gives stellar mass $M_*=3 \times 10^{10} M_{\sun}$, 
SFR=0.007 $M_{\sun}$ yr$^{-1}$
\citep{2015ApJS..219....8C}. }
Its expected $\rm DM_{host}$
are 20 pc cm$^{-3}$ and 28 pc cm$^{-3}$ for the NE2001 and YMW16 models, respectively.
These two sources both have a smaller $\rm DM_{host}$ than FRB 121102. Other host galaxy candidates have even smaller $\rm DM_{host}$ than FRB 121102.

\subsection{FRB 171020}

There are 4974 extended sources within the error box of FRB 171020.
Among them, 31 of them have the redshift information. 
Four of them have spectroscopic redshifts smaller than 0.15, 
and 8 of them have photometric redshifts smaller than 0.15.
The one with the lowest redshift, ESO 601- G 036, 
is the galaxy candidate proposed by \cite{2018ApJ...867L..10M}.
It has $\rm DM_{host}=41$ pc cm$^{-3}$ and 54 pc cm$^{-3}$
for NE2001 and YMW16 models, respectively. For most possible redshifts,
the derived $\rm DM_{host}$ is much smaller than that of FRB 121102. 
Only if the host galaxy is intrinsically very faint (so they are much 
closer) could its $\rm DM_{host}$ reach the lower limit of FRB 121102 $\rm DM_{host}$. In this case, the host galaxy candidate should have an 
absolute magnitude similar to or larger (fainter) than that of FRB 121102.
As shown in Fig. \ref{fig:DMhost_aMag}, 
galaxies without redshift information may achieve 
$\rm DM_{host}=40-60$ pc cm$^{-3}$ if they are extremely nearby.

\subsection{FRB 171213 and FRB 180810.J1159+83}

Both FRB 171213 and FRB 180810.J1159+83 have $\rm DM_{exc} \sim 90$ pc cm$^{-3}$.
It is possible to find a host galaxy candidate similar to 
that of FRB 121102.
Also, they are out of the redshift range for our galaxy-group-based method for the
$z-\rm DM_{IGM}$ relation.
We thus do not explore them in detail.

\subsection{FRB 180814.J0422+73}

FRB 180814.J0422+73 is the second repeating FRB
\citep{2019Natur.566..235C}.
There are only 50 extended sources found in Pan-STARRS, 
and 1 in GLADE in the error box. This is due to the smaller
positional uncertainty compared with other objects. 
The brightest galaxy is the one found in GLADE, with a $g$-band kron magnitude
17.5 mag, and a 2MASS photometric redshift 0.078. 
The second brightest one is a point source with another fainter point source 1.7 arcsec away.
It is quite likely spurious, so
we do not show it in the plot.
Other galaxies are more than one order of magnitude fainter than these two.

There are many galaxy groups near the line of sight of FRB 180814.J0422+73 for $z<0.02$. The host galaxy should have to be very nearby,
if they have a $\rm DM_{host}$ similar to that of FRB 121102.
They should be then intrinsically very faint.
As shown in Fig.\ref{fig:DMhost_aMag}, the host of FRB 180814.J0422+73 has to be
much fainter than $-14$ magnitude, more than 3 orders of magnitude fainter than that of FRB 121102, if a $\rm DM_{host}$ similar to that of FRB 121102 is assumed.
The $\rm DM_{host}$ of FRB 180814.J0422+73 must be very small ($<$7 pc cm$^{-3}$),
if its host is as bright as FRB 121102.
In this case, the galaxy with redshift, 2MASS J042221.4+734710.2, is not the host,
if its photometric redshift is correct.

Even if we use the empirical $z-\rm DM_{IGM}$ relation,
the conclusion is similar.
If 2MASS J042221.4+734710.2 is the host galaxy of FRB 180814.J0422+73, the estimated $\rm DM_{IGM}=67$ pc cm$^{-3}$
\citep{2014ApJ...783L..35D},
indicating $\rm DM_{host} \sim$ 5 pc cm$^{-3}$ DM.
If 2MASS J042221.4+734710.2 is not the host,
the host galaxy should be at least three orders of magnitude fainter than
the host of FRB 121102, that is, $M_{\rm r} > -14$ mag.
For comparison, LMC and SMC have absolute magnitudes $-18.36$ and $-16.82$, respectively. 
In this case, the FRB would be quite local although still extragalactic.
Its isotropic energy would have to be two or three orders of
magnitude smaller than typical FRBs, e.g., FRB 121102. 

In general, we conclude that the host galaxies of nearby FRBs typically have small $\rm DM_{host}$ values, or are intrinsically faint, much fainter than the hosts of LGRBs and SLSNe \citep{2017ApJ...841...14M}. This is in contrast to the conclusion drawn from the FRB 121102 measurement \citep{2017ApJ...834L...7T, 2017ApJ...844...95K} and the statistical analysis of \cite{2017ApJ...839L..25Y}. Future observations of more localized FRBs will test whether a small $\rm DM_{host}$ is typical for nearby FRBs only or for most FRBs in general.

\section{DM contribution from the host galaxies of LGRBs and SLSNe.}

Due to the similarity of the FRB 121102 host with LGRB/SLSNe hosts, 
FRBs are highly believed to be powered by magnetars
born during LGRBs and SLSNe
\citep{2017ApJ...841...14M, 2017ApJ...839L...3K, 2017ApJ...843L...8B, 2017ApJ...843...84N}.
We want to further explore whether the host galaxies of nearby FRBs are similar to those of LGRBs and SLSNe.

Host galaxies of LGRBs and SLSNe are generally star forming dwarf galaxies
\citep{1997Natur.387..476S, 1998ApJ...507L..25B, 2002AJ....123.1111B, 
2002ApJ...566..229C, 2004A&A...425..913C, 2009ApJ...691..182S, 2015A&A...581A.125K}. 
If the galaxy electron density is known, 
the host galaxy DM contribution 
can be estimated based on the scale length $r_{\rm e}$, 
and the offset of the transient from the center of the galaxy $r_{\rm off}$
\citep{2002AJ....123.1111B, 2006Natur.441..463F, 2016ApJ...817..144B, 2016ApJS..227....7L}.

The free electrons in the interstellar medium are generally ionized 
by the death of massive stars. 
They are thus likely correlated to the star formation rate (SFR)
and H$\alpha$ emission \citep{1977ApJ...216..433R}.
We can thus estimate their electron density $n_{\rm e}$ based on their H$\alpha$ emission lines, or SFR.
Since resolved optical emission is not always available, 
we test two possible distributions, i.e. the spherical Gaussian distribution and Milky Way-like distribution. 

We obtained SFR, $r_{\rm e}$, $r_{\rm off}$ and absolute magnitude of SLSNe from \cite{2015ApJ...804...90L}, \cite{2018MNRAS.473.1258S}, and \cite{2016ApJ...830...13P}, 
and those of LGRBs from \cite{2016ApJS..227....7L}.
We then estimate the $\rm DM_{host}$ from LGRBs/SLSNe-like host galaxies
as follows.

\subsection{Spherical Gaussian Distribution}

LGRB and SLSN hosts are dwarf star-forming galaxies, which resemble SMC in many aspects. Following the treatment of SMC by \cite{2017ApJ...835...29Y}, 
we assume that the electron density follows
$$n_{\rm e}=n_0 e^{-(r/r_{\rm e})^2}, $$
where $r_{\rm e}$ is the scale length of the galaxy. 
 
Since LGRBs and SLSNe both highly trace massive stars, 
it is reasonable to assume that they are in the disk plane. 
If the the host is face on, 
for a specific offset $r_{\rm off}$, one has
$${\rm DM} \equiv \int_0^{\infty} n_e dl =\frac{\sqrt{\pi}}{2} n_0 r_e {\rm exp}\left (-\frac{r_{\rm off}^2}{r_{\rm e}^2}\right),$$
$${\rm EM} \equiv \int_{-\infty}^{\infty} n_e^2 dl =\sqrt{\frac{\pi}{2}} n_0^2 r_e {\rm exp}\left (-2\frac{r_{\rm off}^2}{r_{\rm e}^2}\right ),$$
so that 
$${\rm DM}^2=\sqrt{\frac{\pi}{8}}r_e{\rm EM}.$$

According to \cite{1977ApJ...216..433R}, H$\alpha$ surface density is a tracer of EM, i.e.
\begin{eqnarray} 
{\rm EM} & = & 2.75 \left (\frac{T}{10^4K} \right)^{0.9} 
\frac{\Sigma_{\rm H\alpha}}{2.42\times10^{-7} \rm ergs\ s^{-1}\ cm^{-2}\ sr^{-1}} \\
&   & \rm cm^{-6} pc \nonumber \\
& = & 486 \left (\frac{T}{10^4K} \right)^{0.9} 
\frac{\Sigma_{\rm H\alpha}}{10^{-15} \rm ergs\ s^{-1}\ cm^{-2}\ arcsec^{-2}} \\
&   & \rm cm^{-6} pc. \nonumber
\end{eqnarray}
If the H$\alpha$ surface density follows the distribution of EM relative to $r_{\rm off}$, one then has
$$\Sigma_{\rm H\alpha}(r)=\Sigma_{\rm H\alpha0} \exp \left (-2\frac{r_{\rm off}^2}{r_{\rm e}^2} \right),$$
and the H$\alpha$ flux can be written as
$$F_{\rm H\alpha}=\int_0^{\infty} \int_0^{2 \pi} \Sigma_{\rm H\alpha0} \exp \left (-2\frac{r^2}{r_{\rm e}^2}\right ) r{\rm d}r {\rm d} \theta=\frac{\pi r_{\rm e}^2}{2} \Sigma_{\rm H\alpha0}.$$
Combining the relations among DM, EM, $\Sigma_{\rm H\alpha0}$ and $F_{\rm H\alpha}$, one has
\begin{equation}
{\rm DM}^2=486 \times 10^{15} \sqrt{\frac{\pi}{8}} r_{\rm e,pc} \left (\frac{T}{10^4 K}\right )^{0.9}
       \left (\frac{2F_{\rm H\alpha}}{\pi r_{e}^2}\right ) \exp \left  (-2\frac{r_{\rm off}^2}{r_{\rm e}^2}\right ),
\end{equation}
where $F_{\rm H\alpha}$ is in units of ergs s$^{-1}$ cm$^{-2}$,
$r_{\rm e}$ and $r_{\rm off}$ are in units of arcsec,
 $r_{\rm e,pc}$ is in units of pc, and
DM is in units of cm$^{-3}$ pc.

\subsection{Milky Way-like distribution}

We also consider a Milky-Way-like electron density distribution as the template of a disk galaxy
\citep{2017ApJ...835...29Y}, i.e.
\begin{equation}
n_0 \propto \sqrt{\frac{L_{\rm H\alpha}}{r_{\rm e}^3}},
\end{equation}
for each LGRB/SLSNe host galaxy.
The offset is re-scaled by $r'_{\rm off,MW}=\frac{r_{\rm off}}{r_{\rm e}}r_{\rm e,MW}$.
The Milky Way star formation rate, SFR$_{\rm MW}=0.27 M_{\rm \odot} \rm yr^{-1}$
\citep{2015ApJ...806...96L},
and the Milky Way scale length $r_{\rm e,MW}=2.15 \pm 0.14$ kpc
\citep{2013ApJ...779..115B} are used. 
The DM$_{\rm host}$ values estimated with this MW template are presented as blue and grey dots in Fig. \ref{fig:DMhost_aMag}, for SLSNe and LGRBs, respectively.

\subsection{Comparison between the candidates and LGRB/SLSNe hosts}

For both spherical Gaussian distribution 
and MW-like distribution, 
the hosts of LGRBs and SLSNe contribute $\sim$ 100 cm$^{-3}$ pc to DM. 
Only $<$ 10\% of LGRBs have DM$_{\rm host} < $ 30 cm$^{-3}$ pc.  
None of the SLSNe has a DM$_{\rm host} < 30$ cm$^{-3}$ pc. 
They are plotted in Fig. \ref{fig:DMhost_aMag} for comparison.
Blue and grey colors are for SLSNe and LGRBs respectively.
Dots and diamonds are for spherical and MW-like distributions, respectively.

FRB 121102 has an estimated $\rm DM_{host}$ in the range of $55-225$ pc cm$^{-3}$.
It is consistent with the estimated value of LGRBs and SLSNe.
Also, the absolute magnitude of its host galaxy, $-17$, is consistent with
the values for the LGRB and SLSN host samples.

Other FRB host candidates, on the other hand, are not consistent with
the LGRB and SLSN host samples.
The host galaxy candidates for FRB 180729.J1316+55, FRB 171020, FRB 171213, 
and FRB 180814.J0422+73 all have a smaller $\rm DM_{host}$ than the lower 
limit of FRB 121102, 55 pc cm$^{-3}$. 
The host galaxy candidates of FRB 180814.J0422+73 do not overlap with either LGRBs or SLSNe at all. 
The host candidates of FRB 180729.J1316+55 
overlap with 5 of the 37 LGRBs hosts, but no SLSN host.
It is located in the faint, low $\rm DM_{host}$ corner of the LGRB/SLSN host distribution.
FRB 171020, FRB 171213 and FRB180810.J1159+83 pass through
the $\rm DM_{host}$ range of LGRB and SLSN hosts, so the possibility that
their hosts are LGRB/SLSN-like is not ruled out. 
However, All of them are located within the very low end of the LGRB/SLSN $\rm DM_{host}$ distribution.
So, collectively, the
probability that all the nearby FRB hosts are consistent with the 
LGRB/SLSN hosts is extremely low.

\section{FRB 180924 and FRB 190523}

{During the review of this paper, two FRBs are located to 
their host galaxies. 
FRB 180924 is in a massive passive galaxy $z=0.3214$
\citep{2019arXiv190611476B}, 
and FRB 190523 is in a massive galaxy at $z=0.66$
\citep{2019arXiv190701542R}. Both host galaxies are unlike that of FRB 121102, supporting our conclusion that the host of FRB 121102 is atypical. On the other hand, those two host galaxies are also brighter than most of our host candidates. 

In order to apply our method to them, we extend the galaxy group catalog to higher redshifts with \cite{2018MNRAS.475..343W}. 
It covers all sky except the Galactic Plane, extends to redshift 0.4, and has a median redshift 0.24.
479 galaxy groups within it are excluded because they are duplicated with the 2MRS galaxy group.
The electron density and cumulative $\rm DM_{IGM}$ of FRB 180924 and FRB 190523 are also presented in Figure \ref{fig:dmz}.
The derived $\rm DM_{IGM}$ of FRB 180924 at redshift $z=0.3214$ is 121 pc cm$^{-3}$, and that of FRB 180924 at redshift $z=0.66$ is 339 pc cm$^{-3}$.
However, the total halo mass range of galaxy group sample in \cite{2018MNRAS.475..343W} is [$7 \times 10^{13}$, $2 \times 10^{15}$] $M_{\sun}$, much larger than the mass threshold 10$^{12} M_{\sun}$ we applied. As a result, many galaxy groups are likely missed. 
Furthermore, our galaxy groups catalogs do not extend to redshift $z>0.4$, so we are  unable to constrain the $0.4-0.66$ range for FRB 190523.
As a result, the $\rm DM_{IGM}$ obtained with our method should be considered as very loose lower limits for these two FRBs.
By subtracting the $\rm DM_{MW}$ and $\rm DM_{halo}=30$ pc cm$^{-3}$, one gets loose upper limits of $\rm DM_{host}$ for FRB 180924 and FRB 190523:
$\rm DM_{host,FRB180924}<169$ pc cm$^{-3}$ and $\rm DM_{host,FRB190523}<354$ pc cm$^{-3}$, respectively.}

{These two FRBs are also presented in Fig.\ref{fig:DMhost_aMag}. 
The very loose $\rm DM_{host}$ upper limits are also plotted. One can see that most host candidates of nearby FRBs are also much fainter than these two hosts. As these two hosts also differ from that of FRB 121102 in terms of star formation rate and offset between the FRB and the host, one can draw the conclusion that the FRB hosts are very diverse among bursts.}

\section{Conclusion and discussion}

We have searched the host galaxy candidates of nearby FRBs whose $\rm DM_{exc}$ is below 100 pc cm$^{-3}$. 
{Due to the selection criteria, their DM$_{\rm host}$ are expected to be smaller than 100 pc cm$^{-3}$.}
The following conclusions can be drawn:

{1. Not all FRBs reside in environments similar to FRB 121102.
The existence of FRBs with $\rm DM_{host}$ less than the lower limit of FRB 121102 $\rm DM_{host}$ reveals that not all FRBs located in environments similar to FRB 121102. The fact that the hosts of the recently localized FRB 180924 and FRB 190523 are also different from that of FRB 121102 and the host candidates studied in this paper strengthens our conclusion and suggests that FRB hosts are very diverse. }

{2. It is strengthened when we examine the $\rm DM_{host}$ vs $M_{\rm B}$ relation. 
The $\rm DM_{host}$ of FRB180814.J0422+73 must be smaller than 10 pc cm$^{-3}$ if it is a normal galaxy, or it is within a galaxy fainter than -14 magnitude. 
}

2. Based on the required $\rm DM_{host}$ vs $M_{\rm B}$ relation, the host galaxies of FRB 180729.J1316+55, FRB 171020, and FRB 180814.J0422+73
cannot be similar to the hosts of SLSNe, and very likely not similar to the hosts of LGRBs, either. This suggests that LGRBs and SLSNe are likely not the progenitor of most FRB sources.

3. The host galaxies of LGRBs and SLSNe typically contribute to a relatively large $\rm DM_{host} \sim 100$ pc cm$^{-3}$. 

4. We develop an observational galaxy-group-based method to estimate the $\rm DM_{IGM}$ of FRBs. This method can directly address the line-of-sight uncertainty of DM-$z$ relation, even though the results are only reliable up to $z=0.033$ below which the complete galaxy group catalogs are available. Such a method can be applied to infer the distance of other nearby FRBs detected in the future. 

Our results on $\rm DM_{IGM}$ somewhat depends on the assumed density in the intergalactic space, which we discussed in Sec. 3 and Eq. (5).  We have tested the uncertainty by assuming zero electron density for the IGM, in which case we obtain a DM that is smaller by $5-10$ pc cm$^{-3}$. Therefore, our conclusions are not significantly affected by our assumption of the IGM density.  In addition, the result of the cosmological hydrodynamic simulations of galaxy formation with star formation and SN feedback by the GADGET3-Osaka SPH code
\citep{2019MNRAS.484.2632S} suggested a comoving electron density similar to Eq. (5) within a factor of a few.  This also corroborates that the electron density value in Eq. (5) is fairly reasonable. 

\acknowledgments
We thank the referee for the helpful suggestions. 
YL thanks Seunghwan Lim, Huiyuan Wang, Weiwei Xu and Qiang Yuan for helpful discussion.
YL is supported by the KIAA-CAS Fellowship, which is jointly supported by Peking University and Chinese Academy of Sciences. This work is also partially supported by the China Postdoctoral Science Foundation (No. 2018M631242).
KN acknowledges the support by the JSPS KAKENHI Grant Number JP17H01111, and the Kavli IPMU, World Premier Research Center Initiative (WPI).
JJS is supported by the Boya Fellowship.



\end{document}